\documentclass[12pt]{article}

\usepackage{amsmath,amssymb,amsthm,epsf,epsfig,graphicx,latexsym}

\newcommand{\be}{\begin{equation}}
\newcommand{\ee}{\end{equation}}
\newcommand{\bea}{\begin{eqnarray}}
\newcommand{\eea}{\end{eqnarray}}
\newcommand{\bml}{\begin{subequations}}
\newcommand{\eml}{\end{subequations}}
\newcommand{\bfig}{\begin{figure}}
\newcommand{\efig}{\end{figure}}

\begin{document}

\begin{center}
{\large { THE $E(2)$ PARTICLE }} \vskip 2cm
Subir Ghosh\\
\vskip .3cm
Physics and Applied Mathematics Unit,\\
Indian Statistical Institute,\\
203 B. T. Road, Calcutta 700108, India\\
\vskip .3cm
And\\
\vskip .3cm
Probir Pal\\
Physics Department,\\
Uluberia College,\\
Uluberia, Howrah 711315,  India.\\
and\\
S.N. Bose National Centre for Basic Sciences,\\
 JD Block, Sector III, Salt Lake City, \\
Kolkata 700 098, India

\end{center}

with translation invariance) may play a
fundamental role.

\vskip 3cm
 {\bf{Abstract:}}
Recently it has been advocated [1] that for describing nature
within the  minimal symmetry requirement, certain subgroups of
Lorentz group may play a fundamental role.  One such
group is $E(2)$ which induces a Lie algebraic Non-Commutative
spacetime [4] where translation invariance is not fully maintained. We have constructed a consistent structure of Non-commutative phase space for this system and furthermore  we have studied an appropriate  point particle action on it.  Interestingly,  the Einstein dispersion relation $p^2=m^2$ remains intact.
The model is constructed by exploiting a dual canonical phase
space following the scheme developed by us earlier [8].

\vskip 1cm

\noindent {\bf{Introduction and Motivation:}} Symmetry principle
and its realization in nature has played a fundamental role
throughout the development of physics. One of the key players in
this regard is the Poincare group, the connected component of the
Lorentz group along with spacetime translations. Poincare group is
identified as the symmetry of nature by the Special Theory of
Relativity. It is the isometry group of the $3+1$-dimensional
Minkowski space, the arena of relativistic classical and quantum
field theories. However, in recent years, small observed
violations of the discrete symmetries, $C$ (charge conjugation),
$P$ (parity) and $T$ (time reversal) at high energies as well as
the theoretical possibility of violation of Lorentz symmetry,
again at high energy, have given rise to a new paradigm: an
appropriate subgroup of the Lorentz group, together with spacetime
translations  might be sufficient to  explain the (so far)
observed  nature. The underlying criteria is that the  subgroup
together with $C$, $P$ and $T$ will generate the full Lorentz
group. In this scheme, the departures from the discrete symmetries
and (predicted)  Lorentz invariance get connected. This is the
principal idea of Very Special Relativity (VSR) by Cohen and
Glashow \cite{vsr}, (for later works see \cite{vsr1}), who
identify the  subgroups as $T(2)$ (isomorphic to $2$-dimensional
translations), $E(2)$  (isomorphic to $3$-parameter Eucledean
motion), $HOM(2)$, (isomorphic to $3$-parameter orientation
preserving transformations) and lastly $SIM(2)$ (isomorphic to the
$4$-parameter similitude group){\footnote{An echo of this idea is
present in the recently postulated Horava-Lifschitz gravity
\cite{horava} which also gives up the full diffeomorphism symmetry
in a quantum gravity model}}.

In attempting to construct a quantum field theory based on the
above VSR subgroups, Sheikh-Jabbari and Tureanu \cite{jt} noticed
a problem: all the above proper subgroups allow only
one-dimensional representation and hence can not represent the
nature faithfully. However, the authors of \cite{jt} provide an
ingenious resolution of the ``representation problem'': generalize
the normal products of operators as deformed or {\it{twisted}}
coproducts \cite{twist} in terms of which the commutation
relations of the (Lorentz) symmetry generators remain intact along
with their (physically realized) higher dimensional
representation. The reduction in symmetry in going from full
Lorentz group to one of the VSR subgroups can be achieved
\cite{jt} by the ``Drinfeld twist'' \cite{twist}. Finally, a
specific VSR subgroup will induce a specific Drinfeld twist, which
in turn is identified with a particular Non-Commutative (NC)
spacetime structure. The circle is completed by the said VSR
subgroup being the isometry of the NC spacetime or stated simply
the NC structure is invariant under one of the VSR subgroups.

In the present article, our interest lies in studying the behavior
of the simplest non-trivial dynamical system, that of a
relativistic point particle, in an NC spacetimes admitted by one
of the VSR subgroups. We choose the NC spacetime associated with
$E(2)$ \cite{jt} which has a Lie algebraic form of
noncommutativity,
\begin{equation}
 [x^{-},x^i]=ilx^i,
\label{tur}
\end{equation}
where $x^-=(t-x^3)/2~,~i=1,2$ and $l$ is the numerical NC
parameter. It should be mentioned at the outset that strictly
speaking, this NC spacetime does not conform to the VSR principle
since it does not have the full translation invariance. Infact the
only NC spacetime that is allowed by the VSR principle is $[x^\mu
,x^\nu ]=i\theta ^{\mu\nu}$ with constant $\theta ^{\mu\nu}$ as
asserted by \cite{jt}. But we still chose (\ref{tur}) because the
NC structure, being of operatorial (Lie algebraic) form,
construction of a particle model living on it is more interesting
and involved. There are instances where particular NC phase space structures are compatible with  modified
energy-momentum dispersion relations, as for example in the likes
of Doubly Special Relativity (DSR, not to be confused with VSR!)
models \cite{dsr,jk2,dsr1,pal} (in this connection see also \cite{jk1}). Interestingly the present analysis
shows that the  Special Theory of Relativity relation $(p^2=m^2)$
remains intact for $E(2)$ although the result is by no means
obvious. We will closely follow the formalism developed in our
earlier work \cite{pal} since both the $\kappa$-Minkowski NC
spacetime in DSR \cite{dsr,jk2,pal} and the NC spacetime (\ref{tur})
considered here, are Lie algebraic in nature.

At this time let us put our work in the proper perspective. We are working in a classical Hamiltonian framework where the {\it{phase space}} plays the fundamental role and it is imperative that all the Jacobi identities between basic variables are satisfied. In the corresponding quantum theory the latter leads to the associativity property of the operators which is a must. In the present context we will show that  the NC bracket (\ref{tur}) along with validity of Jacobi identity requires  the $\{x^\mu ,p^\nu \}$ bracket to be non-canonical and thus $p^\mu $ no longer behaves as the translation operator. One can try to construct a translation generator but it will not satisfy all the Jacobi identities and so its role in a Hamiltonian scheme is not very clear. However, questions concerning the validity of Jacobi identity becomes much more severe in the quantum theory due to operator ordering ambiguities and regularization problems. Jacobi identity violation and non-associative operators are discussed in quantum mechanics problems  \cite{ano} and quantum field theory problems \cite{anomaly}. Also Jacobi identity constraint can lead to important physical consequences \cite{ano1}.

In recent years High Energy physics in NC spacetime received a
strong impetus after the work of Seiberg and Witten \cite{nc} who
demonstrated that certain low energy limits of open string theory
are dual to an NC gauge theory where the NC parameters are
represented by two form background fields having constant values.
In general, NC spacetimes of the operatorial form, (Snyder
spacetime \cite{sn} being the earliest example), are not very
common mainly because it is difficult to construct a covariant
{\it{phase space}} NC structure keeping in mind the restrictions
imposed by Jacobi identity. The Lie algebraic forms of NC
spacetime ($[x^\mu ,x^\nu ]=iC^{\mu\nu }_\rho x^\rho )$  have
appeared in various contexts such as in fuzzy spheres \cite{fuz},
$\kappa$-Minkowski spacetime in DSR \cite{dsr} and of course  the
new forms given in \cite{jt} for VSR \cite{vsr} (studied in the
present paper) etc. Examples of operator forms of NC phase space
with mixed algebra appear in \cite{masud,ano1,sn}.

Whenever a new NC spacetime is proposed it is a challenging task
to ascertain how the dynamics is affected by the NC nature. This
started with the simplest and most studied NC space, the Moyal
plane, which appeared as an effective configuration space for low
energy (below Landau level) charged particles moving in a plane
with a perpendicular constant magnetic field \cite{nc}. For
obvious reasons, the dynamical model construction based on
operatorial form of NC turns out to be more involved. We
successfully implemented this program in \cite{pal} for the
$\kappa$-Minkowski case where we demonstrated that the particular NC
spacetime is compatible with a modified dispersion (mass-energy) relation
\cite{dsr,ms}.  This is precisely the aim of the present note. We
have constructed a point particle model, valid in the lowest
non-trivial order in $l$, that lives in the NC spacetime
(\ref{tur}). An important observation, as already mentioned, is
that the mass-energy relation remains the  unchanged Einstein
relation.

\noindent {\bf{Formalism:}} Let us quickly recapitulate the scheme
we employed \cite{pal} in constructing relativistic particle
models for  the $\kappa$-Minkowski form in Magueijo-Smolin base
\cite{ms}. It is in principle possible, by way of Darboux's
theorem, to construct an invertible map that takes us from the NC
variables (or phase space) to canonical variables (or phase space)
and back. However, in practice, for specific NC structures it can
be difficult to derive this map explicitly. If one is able to
derive such a map it becomes a very convenient tool to construct
dynamical models in NC phase space simply by starting from a known
form of (canonical) action and then exploiting the map to
reexpress the action in NC variables and subsequently work out the
dynamics. In the $\kappa$-Minkowski case mentioned above the exact
map, (valid to all orders of the NC parameter), was available
\cite{ms,pal} and we used it successfully to construct DSR
particle models.

Returning to the NC spacetime (\ref{tur}), we will follow the same
principle as in \cite{pal}. Unfortunately it is extremely
difficult to construct the {\it{exact}} Darboux map and we
restrict ourselves to the lowest non-trivial order in the NC
parameter $l$.

\noindent {\bf{The present NC model:}} We will not consider the
quantization of the particle model in the present paper. Hence the
commutators are to be interpreted as classical Poisson brackets.
We start by rewriting the NC phase space {\footnote{Some sectors of this algebra are quite close to the ones studied in \cite{jk2} but a closer look reveals that they are distinct in nature.}} in a manifestly covariant
form,
$$
\{x^\mu  ,x^\nu  \} = l[(\eta ^\mu  x^\nu   - \eta ^\nu  x^\mu  )
- ( \alpha ^\mu  x^\nu   - \alpha ^\nu x^\mu ) - ( (\eta x) +
(\alpha x)) ( \eta \,^\mu  \,\alpha \,^\nu  \, - \,\eta ^\nu
\alpha \,^{\mu } )] $$
$$
\{x\,^\mu  \,,\,p\,^\nu  \}  = \, - \,g\,^{\mu \nu }
   - l[ (\eta -\alpha )\,^\mu  \,p\,^\nu  \, + \,
   (\eta \,p)  \alpha \,^\mu  (\eta -\alpha )\,^\nu ] $$
\begin{equation}
\{p^\mu  ,p^\nu  \}  = 0 \label{1}
\end{equation}
For $l=0$ the algebra reduces to the canonical one. We use the
short form $(ab)=a^\mu b_\mu)$. We have introduced two constant
unit vectors, a timelike one $\eta _\mu $ and a spacelike one
$\alpha _\mu $,
\begin{equation}
\eta ^\mu = (1,0,0,0)~;~\alpha ^\mu
 = (0,0,0,1)~;~\eta ^2  = 1~;~\alpha ^2 =- 1~;~\alpha^\mu \eta_\mu =(\alpha \eta )  = 0.
\label{def}
\end{equation}

It is important to note that in (\ref{1}) we have provided the
full NC phase space which did not appear in \cite{jt} or
elsewhere. We have derived (\ref{1}) by demanding validity of
Jacobi identity for all possible combinations of phase space
degrees of freedom. It is quite obvious that the mixed Jacobi
identity for $x^\mu ,x^\nu ,p^\lambda $ will never be satisfied
with a canonical $\{x^\mu ,p^\nu \}=-g^{\mu\nu}$ Poisson bracket.
Note that for the present algebra, all the Jacobi identities are
exactly satisfied. This also means that $p^\mu $ is no longer the correct translation generator. We consider the Jacobi identities to be sacred,
especially as their validity is directly connected with the
associativity of operators upon quantization.

\noindent {\bf{Canonical Variables, Symmetry Generators and
$E(2)$-Invariance of the NC Phase Space Algebra:}}
 In order to compute the Darboux map, corresponding to the NC bracket system (\ref{1}), let us define a set of
 canonical phase space variables $X^\mu,P^\mu$,
 \begin{equation}
\{X^\mu  ,P^\nu  \}  =  -  g^{\mu \nu } \,;\, \{X^\mu  ,X^\nu  \}
=\{P^\mu  ,P^\nu  \} = 0.
 \label{2}
 \end{equation}
After a long computation we obtain the $O(l)$ Darboux map,
 \begin{equation}
p^\mu =P^\mu~;~x^\mu =X^\mu +l[ (\,XP)(\eta  - \,\alpha )^\mu \
  + \,(\eta \,P\,)\,\{(\,\eta X) -(\alpha X )\}\,\alpha \,^\mu ].
 \label{3}
 \end{equation}
The inverse map is the following:
 \begin{equation}
X^{\mu }  = x^\mu   - l[  (\eta p)\{(\eta x) -(\alpha x)\}\alpha ^\mu +
(px)(\eta -\alpha ) ^{\mu }] .
 \label{4}
 \end{equation}

 We follow our earlier prescription \cite{pal} where the Poincare algebra
remains {\it{undeformed}}. Hence the generators are first
constructed in the canonical phase space where they have the
conventional structure and automatically satisfies the canonical
Poincare algebra. This leads us to the Lorentz (rotation)
generator,
$$
J\,^{\mu \,\nu \,}  = \,X\,^{\mu \,} P\,^{\nu \,\,}
  - \,\,X\,^{\nu \,} \,P\,^{\mu \,}$$
\begin{equation}
=[x^\mu   - l\{  (\eta p)((\eta x) -(\alpha x))\alpha ^\mu + (px)(\eta
-\alpha ) ^{\mu })\}]p^\nu -(\mu \rightleftharpoons \nu). \label{5}
\end{equation}
In the second line $J^{\mu\nu }$ is expressed in terms of NC
variables  using the inverse map (\ref{4})  and expectedly it
contains extra $l$-dependent operator terms besides the canonical
structure. With the latter expression one needs to use the NC
brackets (\ref{1}).

It is interesting to note that the transformation of $p^\mu $
still remains canonical,
\begin{equation}
 \{J\,^{\mu \,\nu \,} ,\,p^{\lambda \,} \}
 = \, - \,g\,^{\mu \,\lambda \,} p^{\,\nu \,}
 + \,g\,^{\,\nu \,\lambda \,} p\,^{\mu \,} ~;~~
 \{J\,^{\mu \,\nu \,} ,\,p^{2\,} \}  = \,0\,.
\label{6}
\end{equation}
This observation is crucial because it shows that there is no need
to alter the energy momentum dispersion relation and the Special
Theory dispersion,
\begin{equation}
p^2=m^2 \label{d}
\end{equation}
is Lorentz invariant and can serve equally well in the presently
studied NC space. This is unlike the case of DSR \cite{dsr,ms,pal}
where a modification is needed in the dispersion relation to make
it invariant in the NC case.

However, the transformation rule for $x^\mu $ undergoes drastic
changes,
$$
 \{J^{\mu \nu } ,x^\lambda \} = \,\,( - \,g^{\mu \,\lambda \,} x\,^{\nu \,}+g^{\nu \,\lambda \,} x^\mu  )$$
\begin{equation}
 + l[\{(\eta x) -(\alpha x)\}p^\mu \eta ^\nu \alpha ^\lambda +(\eta
p)x^\mu (\eta -\alpha )^\nu \alpha ^\lambda  -(xp)(\eta -\alpha
)^\mu g^{\nu\lambda } $$$$-(\eta p)\{(\eta ) -(\alpha x)\} \alpha ^\mu
g^{\nu\lambda })]-(\mu \rightleftharpoons \nu). \label{7}
\end{equation}
First of all we need to ensure that the relevant configuration
space variables behave correctly under $J_z \equiv J^{12}$:
\begin{equation}
\{J^{12},x^\lambda \}=x^1g^{2\lambda }-x^2g^{1\lambda }
 \label{rot}
\end{equation}
where we have used (\ref{7}). This shows that $x^i$ behaves as a vector. Also from (\ref{7}),
$\{J^{12},x^0 \}=\{J^{12},x^3 \} =0$ we find that $x^{\pm}$ is
invariant under $J^{12}$. These are same as the transformations
given in \cite{jt}.

Furthermore, in our formalism, the
Poincare generators in (\ref{5}) are expressed in
terms of the canonical variables $(X^\mu , P^\mu )$. Hence, by
construction they will obey the {\it{canonical}} Poincare algebra. This means the the $E(2)$ generators $T_1 , T_2 , J_z$  constructed out of the Poincare generators (see below) will satisfy the canonical $E(2)$ algebra,
$$ \{T_1 , T_2 \} = 0, ~~\{J_z , T_1 \} = −T_2 ,~~ \{J_z , T_2 \} = T_1. $$

We have also seen that under Poincare transformations, with these
generators, the momentum $p^\mu $ transforms canonically (\ref{6})
whereas the position $x^\mu $ fails to do so, as shown in
(\ref{7}).

Before proceeding with specific model building we need to ensure
that, in our prescribed scheme, the NC algebra (\ref{1}) is $E(2)$
invariant. In other words this means that we have to check the
{\it{stability}} of the NC symplectic structure (\ref{1}) under
$E(2)$ transformations. In general, for a set a generators (of
symmetry transformations) $J^a$ and a generic symplectic structure
$\{A,B\}=C$ one needs to check the identity below:
\begin{equation}
\delta ^{J^a} \{A,B\}=\delta^{J^a} C~~;~~~\delta ^{J^a}
A=\{J^a,A\}, \label{st0}
\end{equation}
and explicitly the left hand side of (\ref{st0}) means
\begin{equation}
\delta^{J^a} \{A,B\} = \{J^a,\{A,B\}\}=\{\delta^{J^a} A,B\}+
\{A,\delta^{J^a} B\} \label{sst}
\end{equation}
Now notice that the above equations (\ref{st0},\ref{sst}) can be rewritten as,
\begin{equation}
 \{\{J^a,A\},B\} + \{\{B,J^a\},A\} +\{\{A,B\},J^a\} =0,
\label{j}
\end{equation}
which is nothing but the Jacobi identity concerning the three operators $J^a,A,B$. Recall that our basic phase space variables satisfy all the Jacobi identities. Hence it is clear that the composit operators, constructed out of these basic variables will also obey the Jacobi identities{\footnote{Indeed, violations of Jacobi identities can arise for composite operators in the quantum theory due to operator ordering and regularizing problems (in quantum field theory, see for example \cite{ano,anomaly,ano1}).}}.

Returning to the problem at hand, one can consider the Poincare generators, (translation and Lorentz rotation), in place of $J^a$ and replace $A,B$ by the phase space variables $x^\mu ,p^\nu  $ and the Jacobi identities will be preserved. In the particular case of $E(2)$ the set of generators consists of
$T_1 = K_x + J_y, T_2 = K_y − J_x, J_z $ where $J_i=\frac{1}{2}\epsilon_{ijk}J_{jk}$ and $K_i=J_{0i}$ with $i = x, y, z$ are respectively generators of rotations and boosts.  As we have argued just now, since Jacobi identities individually for $J_{\mu\nu}$ and $P_\mu $ will be satisfied, any combinations of them, in particular the $E(2)$ generators defined above, will also respect the Jacobi identities. Indeed the computations are straightforward and one can explicitly check (which we have done) that the Jacobi identities are satisfied for $E(2)$ generators in place of generic operators $J^a$. This means that the stability of the noncommutative (phase space) bracket structure that we have proposed in (\ref{1})  is stable under $E(2)$ transformations.

The careful reader is probably worried (and rightly so) since it is understood that all  the Poincare symmetry transformations are not valid for VSR. The quick example that is given is that translation symmetry in $i$-directions ($x$ and $y$ directions) are lost since the NC bracket is $\{x^-,x^i\}=lx^i$ as given in (\ref{tur}). This assertation is true provided one considers the translation generator ${\cal{T}}^\mu $ that satisfies $\{ {\cal{T}}^\mu,x^\nu \}=g^{\mu\nu}$ between the translation generators $(Tr)^\mu$ and $x^\mu $. However, as we have emphasized before, (see below (\ref{def})), it is not possible to satisfy {\it{all}} the Jacobi identities between $x^\mu , {\cal{T}}^\mu $ with $\{x^\mu ,x^\nu \}$ being noncommutative but $\{ {\cal{T}}^\mu,x^\nu \}=g^{\mu\nu}$ being canonical.  On the other hand,  it is enough, (at least for the present purpose), to have the NC  algebra $\{x^-,x^i\}=lx^i$ as a necessary property of the spacetime and validity of all the phase space Jacobi identities as a necessary requirement to construct a consistent point particle action, as shown in the next section. It is important to point out that in the Hamiltonian framework,  the full phase space has to be considered. If one was concerned only with the coordinate space then, by itself, the coordinate Jacobi identity $\{x^\mu ,\{x^\nu ,x^\lambda \}\} +~cyclic~terms =0$ even with the NC bracket $\{x^-,x^i\}=lx^i$ as given in (\ref{tur}). The need to modify the $\{x^\mu ,p^\nu \}$ bracket as in (\ref{1}) comes when one has to satisfy the mixed Jacobi identity $\{x^\mu ,\{x^\nu ,p^\lambda \}\} +~cyclic~terms =0$.

 Thus finally we have the important result that the NC
symplectic structure proposed in \cite{jt}  transforms
{\it{covariantly}} under the generators of
$E(2)$. We stress that there is no clash between our formalism and the idea of VSR (regarding it not being invariant under all translations) simply because the momenta $p_\mu $ in (\ref{1}) are not the translation generators because of the additional terms in the $\{x^\mu ,p^\nu \}$ commutator in (\ref{1}). Indeed, for the translation generators ${\cal{T}}^i$ the NC bracket (\ref{1}) ceases to be invariant. \\
\noindent {\bf{Point Particle Model and First Order Action:}} Now
we are faced with the problem of writing a suitable action in the
NC phase space that will be consistent with the NC brackets
(\ref{1}) and correct dispersion relation (\ref{d}). It becomes
 clear at once how useful the Darboux construction is because as in
the case of Lorentz generators (\ref{5}) we can once again start
with well-known relativistic canonical action for a massive
particle,
\begin{equation}
L\, = \dot X\,^\mu  \,P^{\mu \,} \, - \,\,\frac{\lambda }{2}
 \,(\,P^{2\,}  - \,m\,^{2\,} ).
\label{8}
\end{equation}
(The procedure is same as that followed by us in \cite{pal}.) Now
we must go over to the NC variables (via the inverse map
(\ref{4})) thus obtaining the cherished form of the action,
\begin{equation}
L\, = (\,p\,\dot x\,)\, -\,l\,[\{(\eta p) -(\alpha p)\}(xp \dot )+(\eta p)(\alpha p)\{(\dot x\eta ) -(\dot x \alpha )\}$$$$+\{(\eta x)-(\alpha x)\}(\alpha p)(\eta \dot p)]-\frac{\lambda }{2}(p^2 -m^2).
\label{9}
\end{equation}
This is a first order action with a modified symplectic structure
but un-modified dispersion relation. In order to check whether the
NC phase space (\ref{1}) is recovered it is straightforward to
perform the Hamiltonian constraint analysis of Dirac \cite{dir},
the saliant points of which are briefly described below.

\noindent {\bf{Dirac Constraint Analysis:}} In the Hamiltonian
formulation of constrained system \cite{dir} any relation between
dynamical variables, not involving time derivative is considered
as a constraint. Constraints can appear from the construction of
the canonically conjugate momenta (known as Primary constraint) or
they can appear from demanding time invariance of the constraints
(Secondary constraint).

Once the full set of constraints is in hand they are classified as
First Class Constraint (FCC) or Second Class Constraint (SCC)
according to whether the constraint Poisson bracket algebra is
closed or not, respectively. Presence of constraints indicate a
redundancy of Degrees Of Freedom (DOF) that is not all the DOFs
are independent. FCCs signal local gauge invariances in the
system. If FCCs are present, there are two ways of dealing with
them (in the quantum case). Either one keeps all the DOFs but
imposes the FCCs by restricting the set of physical states to
those satisfying $(FCC)\mid state >=0$. On the other hand one is
allowed to choose further constraints, known as gauge fixing
conditions so that these together with the FCCs turn in to an SCC
set and these will also give rise to Dirac brackets that we
presently discuss. In case of SCCs, say for $SCC_1,SCC_2$ with
$[SCC_1,SCC_2]=c$ where $c$ is not another constraint, to proceed
as before with $(SCC)\mid state >=0$ one reaches an inconsistency
because in $<state \mid[SCC_1,SCC_2]\mid state >=<state \mid c
\mid state
>$ the $LHS=0$ but $RHS\neq 0$. For consistent imposition of the SCCs one defines the Dirac
brackets between two generic variables $A$ and $B$,
\begin{equation}
\{A,B\}_{DB}=
\{A,B\}-\{A,SCC_i\}\{SCC_i,SCC_j\}^{-1}\{SCC_j,B\},\label{di}
\end{equation}
where $SCC_i$ are a set of SCC and $\{SCC_i,SCC_j\}$ is the
constraint matrix. For SCCs this matrix is invertible and since
$\{A,SCC_i\}_{DB}=\{SCC_i,A\}_{DB}=0$ for all $A$ one can
implement $SCC_i=0$ strongly meaning that some of the DOFs can be
removed thereby reducing the number of DOFs in  the system but one
must use the Dirac brackets. Hence, SCCs induce a change in the
symplectic structure and subsequently one quantizes the Dirac
brackets. Same principle is valid for gauge fixed FCC system
mentioned before.

\noindent {\bf{Recovering the NC Phase Space:}} In the present
model (\ref{9}), canonical momenta for $x^\mu $ and $p^\mu $,
$$\pi ^\mu _x=\frac{\partial L}{\partial \dot x_\mu}~;~~\pi ^\mu _p=\frac{\partial L}{\partial \dot
p_\mu},$$ yields the constraints,
\begin{equation}
\psi _{1\,}^\mu   = \,\pi _{x\,}^\mu  \, - \,p^{\mu \,}
 + \,l\,\,(\eta p -\alpha \,p )\,p^{\mu \,}
 + \,l\,(\alpha p\,)\,(\,\eta \,p\,)\,(\eta -\alpha )\,^{\mu
 \,} \approx 0,
\label{10}
\end{equation}
\begin{equation}
\psi _{2\,}^\mu  \, = \,\pi _{p\,}^\mu   +\,l\,\{(\eta p) -(\alpha \,p )\}x^{\mu \,}
 + \,l\,(\alpha p\,)\{(\eta x)-(\alpha x) \}\eta ^{\mu
 \,}\approx 0.
\label{11}
\end{equation}
The set of constraints $\psi _a^\mu ,a=1,2$ turn out to be SCC and
the constraint bracket matrix $\{\psi _a^\mu ,\psi _b^\nu \}$ is
computed below:

\begin{equation}
\{ \psi_a^\mu,\psi_b^\nu\}=
 \left (
\begin{array}{cc}
 0 & B^{\mu\nu} \\
-B^{\nu\mu} & C^{\mu\nu}
 \end{array}\right )\label{mat1}
\end{equation}
$$ B^{\mu\nu}=
   g^{\mu \nu } \, - l\,(\eta -\alpha ) ^\nu  p ^\mu   -
   l(\eta p)(\eta -  \alpha ) ^\mu \alpha ^\nu ,   $$
$$ C^{\mu\nu}=
   l(( \eta -\alpha )^\mu  x^\nu   - ( \eta -\alpha )^\nu  x^\mu +
   l\{(\eta x)-(\alpha x)\}(\alpha ^\mu  \eta ^\nu   - \alpha ^\nu  \eta ^\mu ).  $$
The non-vanishing inverse of the above  matrix is,
\begin{equation}
(\{ \psi^\nu,\psi^\rho\})^{-1}_{ab}=
 \left (
\begin{array}{cc}
 D^{\mu\nu} & E^{\mu\nu} \\
-E^{\nu\mu} & 0
 \end{array}\right )\label{mat2}
\end{equation}
$$ D^{\mu\nu} =l((\eta -\alpha )^\mu x^\nu -(\eta -\alpha )^\nu x^\mu +\{(\eta x) -(\alpha x)\}(\alpha ^\mu \eta ^\nu -\alpha ^\nu \eta ^\mu ),$$
$$ E^{\mu\nu}=- g^{\mu\nu }  - l(\eta -\alpha ) ^\mu p^\nu -l(\eta p)(\eta -\alpha ) ^\nu \alpha ^\nu .$$
Using the constraints (\ref{10},\ref{11}), inverse of the
constraint matrix (\ref{mat2}) and the definition of the Dirac
brackets (\ref{di}), it is easy to convince oneself that indeed,
the NC phase space algebra (\ref{1}) is recovered.

\noindent {\bf{Namu-Goto Action:}} The remaining task is to
construct a Nambu-Goto like point particle action for the present
model. This means that we need to eliminate the momenta $p^\mu $
and the multiplier field $\lambda $ from the first order action.
To that end, we first compute the variational equations of motion
from the action,
\begin{equation}
\dot x_\mu +l[\eta _\mu \{\{(\eta x)-(\alpha x)\}(\alpha \dot p)-(xp \dot)\}+\alpha _\mu  \{(xp\dot )-(\eta p)((\eta \dot x)-(\alpha \dot x))$$$$ - (\eta \dot p)((\eta x) -(\alpha x))\}+x_\mu\{(\eta \dot p)-(\alpha \dot p)\}]=\lambda p_\mu ,
\label{v1}
\end{equation}

\begin{equation}
\dot p_\mu -l[(\eta p)(\alpha \dot p )(\eta _\mu -\alpha _\mu )+\{(\eta \dot p)-(\alpha \dot p )\}p_\mu ]=0.
\label{v2}
\end{equation}
Rewriting (\ref{v2}) as $\dot p_\sigma G^{\sigma \mu }=0$ one can check the to $O(l)$ the inverse of $G^{\sigma \mu }$ exists and so $\dot p_\mu =0$ is a consistent solution. Putting this back in (\ref{v1}) and after squaring we obtain,
\begin{equation}
\lambda =\frac{\sqrt{\dot x^2}}{m}-l[\frac{(\eta \dot x)(\alpha \dot x)\{(\eta \dot x)-(\alpha \dot x)\}}{\dot x^2}+\{(\eta \dot x)-(\alpha \dot x)\}],
\label{v3}
\end{equation}
where we have exploited the mass-shell constraint and considered only terms of $O(l)$. To the same order we also obtain
\begin{equation}
(p\dot x)=m\sqrt{\dot x^2}.
\label{v4}
\end{equation}
Collecting all the terms we obtain the cherished Nambu-Goto form of the Lagrangian for the $E(2)$ particle to order $l$:
\begin{equation}
L=m\sqrt{\dot x^2}-lm^2 \{(\eta \dot x)-(\alpha \dot x )\}\{1+\frac{(\eta \dot x)(\alpha \dot x )}{\dot x^2}\}.
\label{v5}
\end{equation}
Clearly for $l=0$ the action reduces to the conventional
Nambu-Goto action for a relativistic massive particle. The first
order action (\ref{9}) as well as its (classically) equivalent
Nambu-Goto form (\ref{v5}) are major results of the present work.\\
\noindent {\bf{Recovering the NC phase Space (Once Again):}} It is
an interesting problem to analyze this particular Nambu-Goto
Lagrangian (\ref{v5}) given its involved time-derivative structure.
This will act as an internal consistency check as well. Note that
although formally it does not contain higher time-derivative
terms, the peculiar nature of the terms forces us to treat this
model as a higher derivative one. We will follow the prescription
of \cite{luk} where one replaces selectively some time derivatives
by other new auxiliary degrees of freedom (thus rendering the
system with less number of time derivatives) and imposes new
constraints so that the model physically remains unchanged.

Let us define $\dot x_\mu = p_\mu $ and rewrite the Lagrangian
accordingly,
$$
L = \Pi ^{(x)} _\mu (\dot x_\mu  - p_\mu ) + \Pi ^{(p)} _\mu
p_\mu  - \frac{\lambda }{2}(\Pi ^{(p)2}   - m^2 ) $$
\begin{equation}
  - l[ (\alpha p)(\eta p))\{(\eta \dot x)-(\alpha \dot x )\} $$$$+ (xp \dot)\{(\alpha p)-(\eta p)\}+(\eta \dot p)(\alpha p)\{(\eta x)-(\alpha x)\}].
\label{18}
\end{equation}
Notice that $\Pi ^{(x)} _\mu $ acts as a Lagrange multiplier
enforcing the above identification as a constraint. Integrating
out $\lambda $ and $\Pi ^{(p)} _\mu$ from the Lagrangian equations
of motion we find,
\begin{equation}
p_\mu  - \lambda \Pi ^{(p)} _\mu  = 0 ~;~~\lambda  = \frac{{\sqrt
{p^2 } }}{m}~;~~\Pi ^{(p)} _\mu  = \frac{{mp_\mu }}{{\sqrt {p^2 }
}}. \label{19}
\end{equation}
Substituting these back in (\ref{19}) in the $\lambda =1 $ gauge
one recovers the Nambu-Goto Lagrangian (\ref{v5}) showing the
equivalence between the Nambu-Goto and first order form. This
demonstration is relevant because in the formalism \cite{luk}
applied here it is clear that construction of the first order form
from the higher order form is not unique. It is expected that if
one chooses different ways in reducing the higher order form to
first order form the resulting actions including the constraint
structure will be different but there will be explicit relations
connecting the sets of variables.

Constraints obtained from the first order action,
\begin{equation}
\Psi ^{(1)} _\mu  = P^{(x)} _\mu  - \Pi ^{(x)} _\mu  + l[(\eta p)(\alpha p)(\eta _\mu -\alpha _\mu )+\{(\eta p)-(\alpha p)\}p_\mu ],
\label{20}
\end{equation}

\begin{equation}
\Psi ^{(2)} _\mu  = P^{(p)} _\mu  +l[\{(\eta p)-(\alpha p)\}x_\mu +\{(\eta x)-(\alpha x)\}(\alpha p)\eta_\mu ].
\label{21}
\end{equation}
where $P^{(x)} _\mu  = \frac{{\partial L}}{{\partial \dot
x_\mu}}~,~P^{(p)} _\mu  = \frac{{\partial L}}{{\partial \dot p_\mu
}}$ are identical to the ones obtained previously
(\ref{10},\ref{11}) ensuring that the same NC phase space
(\ref{1}) will reappear.

\noindent {\bf{Conclusion:}}  In this paper we have focussed on a
particular VSR subgroup $E(2)$ of the Lorentz group and  the
induced    Non-Commutative spacetime with a Lie algebraic form
\cite{jt}. We have constructed a relativistic point particle model
that lives in this Non-Commutative phase space and enjoys an
undistorted energy-momentum dispersion relation. Our results are
restricted to the first non-trivial order in $l$ - the
non-commutative parameter.

An important task that remains is to extend the model to all
orders in $l$. In principle this should be possible due to the
Darboux theorem. Another interesting area will be to study the
solutions of the equations of motion for the ``free'' $E(2)$
particle. (Actually the single particle theory may not be free in
the conventional sense.) It might also be possible to introduce
external gauge interactions by way of minimal couplings.$\dot )$

\vskip .4cm \noindent {\bf{Acknowledgment:}} It is a pleasure to
thank Anca Tureanu for helpful correspondence at early stage of our
work.

\end{document}